\documentclass{ws-ijmpe}
\usepackage[super,compress]{cite}
\usepackage{graphicx}
\usepackage{amsmath, amssymb}

\begin{document}

\markboth{V. N. Tarasov et al.}{Stability Peninsulas on the Neutron Drip Line}

%
\catchline{}{}{}{}{}
%

\title{LIGHT EXOTIC NUCLEI WITH EXTREME NEUTRON EXCESS AND $2 \leq Z \leq 8$}

\author{\footnotesize V.N.~TARASOV$^{1}$, K.~A.~GRIDNEV$^{2,3}$, 
V.I.~KUPRIKOV$^{1}$, D.~K.~GRIDNEV$^{2,3}$, D.V.~TARASOV$^{1}$, K.S.~GODBEY$^4$, 
X.~VI{\~N}AS$^{5}$ and WALTER GREINER$^3$}

\address{$^1$NSC, Kharkov Institute of Physics and Technology, Kharkov, 61108 
Ukraine}
\address{$^{2}$Institute of Physics, St. Petersburg State University, St. 
Petersburg, 198504 Russia}
\address{$^3$Frankfurt Institute of Advanced Studies, Frankfurt am Main, 60438 
Germany}
\address{$^4$Vanderbilt University, Nashville, Tennessee 37235, USA}
\address{$^5$University of Barcelona, Barcelona, E-08028 Spain}

\maketitle

\begin{history}
\received{(received date)} \revised{(revised date)}
\end{history}

\begin{abstract}
Using HF+BCS method we study light nuclei with nuclear charge in the range $2 
\leq Z \leq 8$ and lying near the neutron drip line. The HF method uses 
effective Skyrme forces and allows for axial deformations. We find that the 
neutron drip line forms stability peninsulas at $^{18}$He and $^{40}$C. These 
isotopes are found to be stable against one neutron emission and possess the 
highest known neutron to proton ratio in stable nuclei. 
\end{abstract}



\section{Introduction}
\label{intro}
Nuclei with extreme neutron access lying close to the neutron drip line 
constantly attract interest both in theoretical and experimental research. 
Experiments with radioactive nuclear beams that are conducted in Dubna, GSI, 
RIKEN have opened new opportunities for obtaining exotic nuclei with impressive 
neutron excess. The construction of FRIB in Michigan \cite{frib} (the 
operational run is planned in 2022) would boost experimental abilities to 
approach the neutron drip line.  
Among major fundamental microscopic approaches in the studies of nuclei with 
neutron excess one can name the Hartree-Fock-Bogoliubov approach (HFB),  
Hartree-Fock plus BCS pairing (HF+BCS) with effective forces and relativistic mean field theory \cite{1,2,3,4,5,6}. 

In Refs.~\refcite{7,8,9,10,11,12,13,14} we explored the possibility of existence of 
islands and peninsulas of stability of nuclei with large neutron excess lying 
beyond the conventional neutron drip line. The calculations in these papers were 
done with the HF method using various type of effective Skyrme forces 
\cite{15,16,17,18,19} and accounting for the axial deformation; the pairing was 
treated in the BCS scheme. We have demonstrated that in the regions of the 
nuclear chart corresponding to extreme neutrons excess around magic and ``new'' 
magic numbers N = 32, 58, 82, 126, 184, 258 there may exist peninsulas of 
neutron stability stretching beyond the conventional neutron drip line. In this 
picture a neutron rich nucleus, which is unstable against neutron separation, 
may regain stability if one adds to it a certain number of neutrons thereby 
shifting it to the stability peninsula. In the HF approach this stability 
restoration is a result of complete filling of neutron subshells with large 
angular momentum. These subshells have a large centrifugal barrier and being 
partially filled they are located in the continuous spectrum, which makes the 
corresponding nuclei unstable against neutron emission. However, when the 
neutron number increases these subshells descend into discrete spectrum and 
nuclei regain stability. The main aim of the present paper is to find stable 
nuclei with the highest possible neutron to proton ratio N / Z, which we believe 
sets a theoretical upper bound for this ratio. We shall consider only nuclei 
with $2 \leq Z\leq 8$, since it has been shown earlier \cite{10,11,12}, the 
maximal N / Z ratio can be attained only for light nuclei. The largest values of 
$N$ are obtained at stability peninsulas. The most extended stability peninsulas 
are obtained with the SkI2 \cite{11,12,13} set of Skyrme forces, therefore, we shall 
use mostly the SkI Skyrme parameters. Let us note that various types of SkI 
forces were obtained and effectively used in Ref.~\refcite{18}. The same type of forces 
was used in Ref.~\refcite{20} for the description of light exotic nuclei, however, 
without accounting for deformation, which is, certainly, a disadvantage 
compromising the predictive power of the obtained results. 

\section{Methods and Results}

The detailed exposition of the method of solving the deformed Hartree-Fock (DHF) 
equations is given in Refs.~\refcite{8,21,22,23}. Here we present the results of the 
calculations using SkI2 set of Skyrme forces. 
One particle wave functions in DHF are expanded in the axially deformed harmonic 
oscillator basis. The principle quantum number for the harmonic oscillator basis 
usually does not exceed $N_0$ = 18 (amounting to 1330 basis functions). This 
dimension of the basis is more than necessary for the calculations of the 
isotopes with $2 \leq Z \leq 8$, which provides high accuracy and reliability of 
obtained results. 

In case of nuclei lying on stability peninsulas along with the DHF calculations 
we do additional calculations, where HF equations are solved directly in 
coordinate space under assumption of spherical symmetry (SHF method) \cite{24}. 
This is reasonable because nuclei belonging to stability peninsulas indeed 
possess spherical symmetry \cite{7,8,9,10,11,12,13,14}. SHF calculations make 
possible the analysis of the role of continuum states in the cases when the DHF 
solution is spherically symmetric. 
We used the BCS constant pairing, where the pairing constant is set equal to $G= 
(19.5/(N+Z))[1 \pm 0.51(N-Z)/(N+Z)]$ \cite{25}, ``+'' and  ``-'' correspond to 
protons  and neutrons respectively. In DHF calculations pairing is restricted to 
bound one-particle states. This choice of pairing as well as the role of 
continuum states in the structure of nuclei lying at stability peninsulas is 
discussed in Refs.~\refcite{10,11,12,13,14}. Let us note that the inclusion of continuum 
states into present SHF calculations of nuclei with $2 \leq Z \leq 8$, which 
form stability peninsula, did not affect the results that were obtained without 
inclusion of continuum states. This is, however, typical for magic and 
quasi-magic nuclei. Similar behavior has been observed for stability peninsulas 
in other parts of the nuclear chart \cite{12}. 

Fig. 1 shows nuclear chart, where the squares represent nuclei that are stable 
against one neutron emission in DHF calculations with SkI2 forces. Exactly the 
same picture is obtained with SkI1 forces. Grey squares on this chart are 
experimentally known stable nuclei. We determine one neutron drip line from the 
condition $S_n = 0$, where $S_n0$ denotes one neutron separation energy. Thereby, any positive value of $S_n$ means stability 
against one neutron emission, even if $S_n$ is marginally small. We calculate 
one neutron separation energies assuming the validity of approximation made in 
Koopman’s theorem. For given $Z$ the position of the neutron drip line is 
determined as follows. We start with a stable isotope and increase the neutron 
number until stability is lost and one neutron separation energy changes its 
sign. Knowing that one-neutron drip line can exhibit nonmonotonic behavior 
\cite{12} we continue increasing $N$. If at some larger value of $N$ the value 
of $S_n$ changes  its sign again then this means that we have located a 
stability peninsula. 
As we have mentioned such effect of stability restoration appears when neutron 
subshells with large angular momentum getting fully filled intrude from 
continuous spectrum into the bound spectrum \cite{10,11,12,13,14}. (In 
\cite{nature} the formation of stability peninsulas was alternatively termed 
irregular behavior of the neutron drip line). Using SkI2 set of Skyrme forces we 
have found one stability peninsula situated at $^{18}$He with the fully filled 
subshell 1d$_{5/2}$  and another one at $^{40}$C with the fully filled subshell 
1f$_{7/2}$. These isotopes that disrupt the monotonic behavior of the drip line 
are illustrated in Fig. 1, where they are highlighted with red (color online) 
color and have arrows pointing at them. Both of these nuclei possess spherical 
symmetry, which is typical for nuclei with fully filled shells. The neutron to 
proton ratio in $^{18}$He and $^{40}$C is $N / Z = 8$ and $N / Z \simeq 5.67$ 
respectively, which is substantially larger than the so far known neutron to 
proton ratios for stable nuclei  \cite{8,9,10,11,12,13,14}. 
Fig.1 also shows the stability peninsula consisting of one isotope $^{32}$C. 
Later we shall discuss in detail the stability restoration for this isotope. 

Fig. 2 shows one neutron separation energies as functions of $A=(Z+N)$ obtained 
with SkI2 Skyrme forces. One can clearly see $S_n$  changing sign at $^{18}$He, 
which results from completely filled subshell 1d$_{5/2}$ being bound. One 
neutron separation energy of this isotope as estimated according to the 
Koopman’s theorem equals 0.45 MeV. Let us consider HF potentials corresponding to the state 1d$_{5/2}$, where this 
state is weakly bound. Because the nucleus is spherical we can use SHF method, 
which has the advantage of representing potentials and wave functions directly 
in coordinate space \cite{10,24}. The DHF  value $S_n$ = 0.450 is close to the 
SHF value $S_n$ = 0.406 MeV. Fig. 3 shows the Fermi level 1d$_{5/2}$ for 
$^{18}$He and the corresponding SHF potential. The dotted line depicts the wave 
function of the Fermi level, which is localized under the centrifugal barrier. 
The height of the centrifugal barrier in the HF potential equals 1.35 MeV, which 
should  additionally enhance the stability of $^{18}$He against one neutrons 
emission. 
When we apply the BCS scheme in case of SHF calculations we also include those localized 
quasibound states with positive energy, whose wave functions are localized under 
the centrifugal barrier. Other continuum states are not considered. In case of 
$^{18}$He and $^{40}$C (SkI2 forces) we considered all localized quasibound 
states. The states were discretized by introducing  boundary conditions, which 
make wave functions vanish outside the sphere with the radius 40 fm. We found 
that the pairing remained zero even when we included localized quasibound states. 

Fig. 4 shows one neutron separation energies for carbon calculated with SkI2 
forces using the DHF method. One can see that $S_n$ increases for $A= 26$, which 
signals the magicity of the number $N = 20$ near the drip line. The neutron 
subshell 1f$_{7/2}$ submerses into bound spectrum becoming completely filled for 
$N=34$, which provides stability against one neutron emission for $^{40}$C. At 
this point $S_n =  0.597$ MeV in DHF calculations and $S_n =  0.594$ MeV in SHF 
calculations. $^{40}$C is a spherically symmetric nucleus. Similar to the case 
of $^{18}$He in Fig. 5 we consider the SHF potential corresponding to the Fermi 
level 1f$_{7/2}$ for $^{40}$C.  The SHF potential for this state has a 
centrifugal barrier with the height 2.48 MeV, which additionally enhances the 
stability of this isotope. In Fig. 4 one witnesses the stability restoration for 
$^{32}$C (26 neutrons) , which is again connected with the state 1f$_{7/2}$. 
Unlike $^{40}$C the nucleus $^{32}$C has quadrupole deformation $\beta = - 
0.32$, and its last filled neutron level has quantum numbers $\Omega = 7/2^{-}$. 
This level, which is the member of the multiplet 1f$_{7/2}$, intrudes into the 
bound part of the spectrum when $N=26$. 

Let us consider in more detail the isotope chains of He and C, which pass through 
$^{18}$He and $^{40}$C. In Figs. 6, 7 one can see the neutron and proton root 
mean square (rms) radii $\langle r^2_{n,p}\rangle^{1/2}$ for helium and carbon. 
The arrows indicate the values for $^{18}$He and $^{40}$C, which form stability 
peninsulas at $N = 16$ and $ N = 34$. Generally in neutron rich nuclei the 
difference of neutron and proton rms radii $\Delta R_{n,p} = \langle 
r^2_{n}\rangle^{1/2}  - \langle r^2_{p}\rangle^{1/2} $ may take large values and 
one can speak about neutron skin or neutron halo effects when $\Delta R_{n,p} > 
0.2$ fm \cite{26}. From the figure one can see large values of $\Delta R_{n,p} $ 
for He and C isotopes in particular  for $^{18}$He and $^{40}$C, which is a manifestation of 
a large neutron halo in these nuclei. In DHF calculations for $^{18}$He we 
obtained $\Delta R_{n,p} = 4.288-2.150 = 2.138$ fm and  $\langle 
r^2_{n}\rangle^{1/2} / \langle r^2_{p}\rangle^{1/2}\simeq 1.994 $. For $^{40}$C 
we got $\Delta R_{n,p} = 4.709-2.800 = 1.909$ fm and  $\langle 
r^2_{n}\rangle^{1/2} / \langle r^2_{p}\rangle^{1/2}\simeq 1.682$. It is useful 
to compare these values with those of extremely neutron rich $^{40}$O \cite{27}, 
where $\Delta R_{n,p} = 1.29$ fm and $ \langle r^2_{n}\rangle^{1/2} / \langle 
r^2_{p}\rangle^{1/2}\simeq 1.436$ (SkM$^*$ forces). This suggests that there is 
a very large neutron halo  in $^{18}$He and $^{40}$C in the picture 
corresponding to SkI2 forces. 

Figs. 8, 9 show neutron and proton density distributions of $^{18}$He and 
$^{40}$C calculated within SHF approximation using SkI2 forces. SHF 
approximation yields the wave function in coordinate representation and is thus 
more appropriate for graphical purposes. Its use in this case is justified by 
the spherical form of these nuclei. Neutron halo in neutron rich nuclei is first 
of all characterized by the spatially extended neutron density distribution 
dominating over the ``normal'' proton density distribution. In this sense Figs. 
8, 9 provide a more clear picture of neutron halo in nuclei forming stability 
peninsulas. Neutron and proton density distributions in Figs. 8, 9 are typical 
for neutron rich nuclei \cite{6}. Our analysis shows that the tail of neutron 
distribution results from weakly bound one particle states. 
\section{Conclusion}

We did HF + BCS calculations of neutron rich nuclei with charges $2 \leq Z \leq 
8$. We have demonstrated that beyond conventional neutron drip line there may 
exist stability peninsulas formed by $^{18}$He and $^{40}$C. These nuclei set 
the record for the largest so far known neutron to proton ratio for stable 
nuclei. They are spherical in shape and have a spatially extended  neutron halo. 
These newly found stability peninsulas complement our previous findings and 
their existence agrees with the general nature of the neutron drip line forming 
peninsulas at magic and quasimagic neutron numbers.


\newpage 

\begin{figure}[th]
\centering
\resizebox{1\textwidth}{!}{%
  \includegraphics{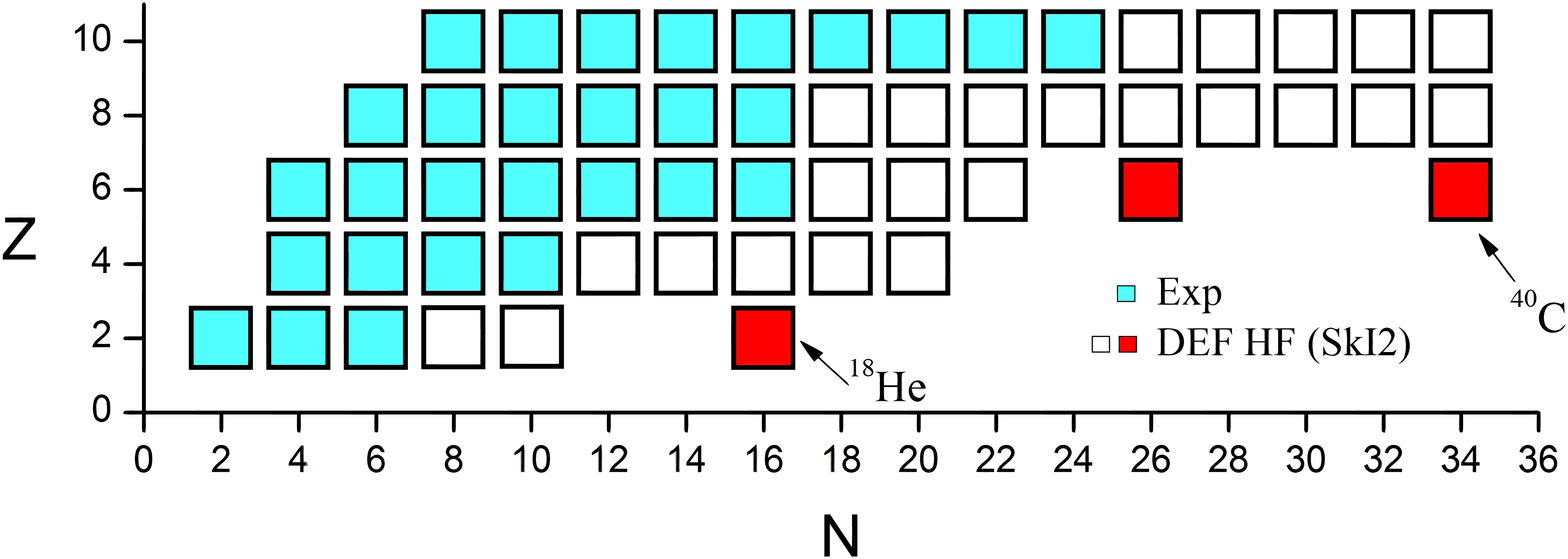}
}
\caption{(Color online). Fragment of NZ chart of even-even nuclei. Empty squares 
are nuclei stable against one neutron emission in DHF calculations with SkI2 
forces. Grey squares represent experimentally known stable nuclei. Black squares 
are stability peninsulas. }
\end{figure}

\begin{figure}[th]
\centering
\resizebox{1\textwidth}{!}{%
  \includegraphics{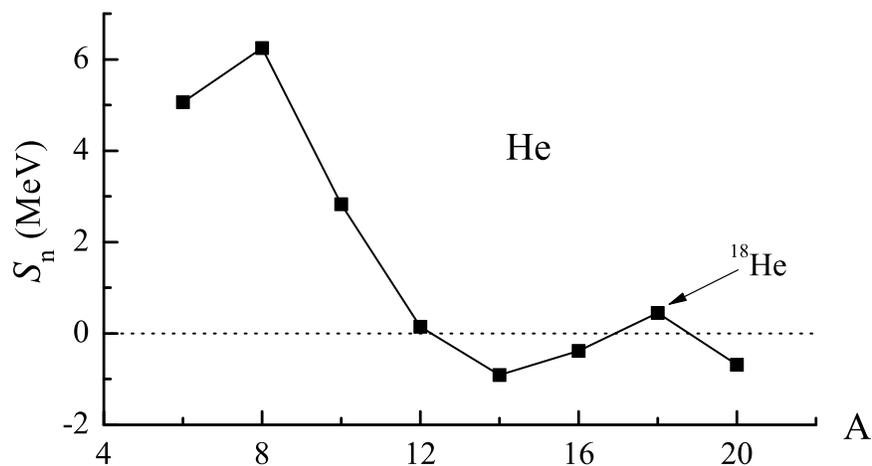}
}
\caption{One neutron separation energies $S_n$ of helium isotopes as a function 
of the neutron number obtained in DHF calculations with SkI2 forces. The arrow 
points at  $^{18}$He. }
\end{figure}

\begin{figure}[th]
\centering
\resizebox{1\textwidth}{!}{%
  \includegraphics{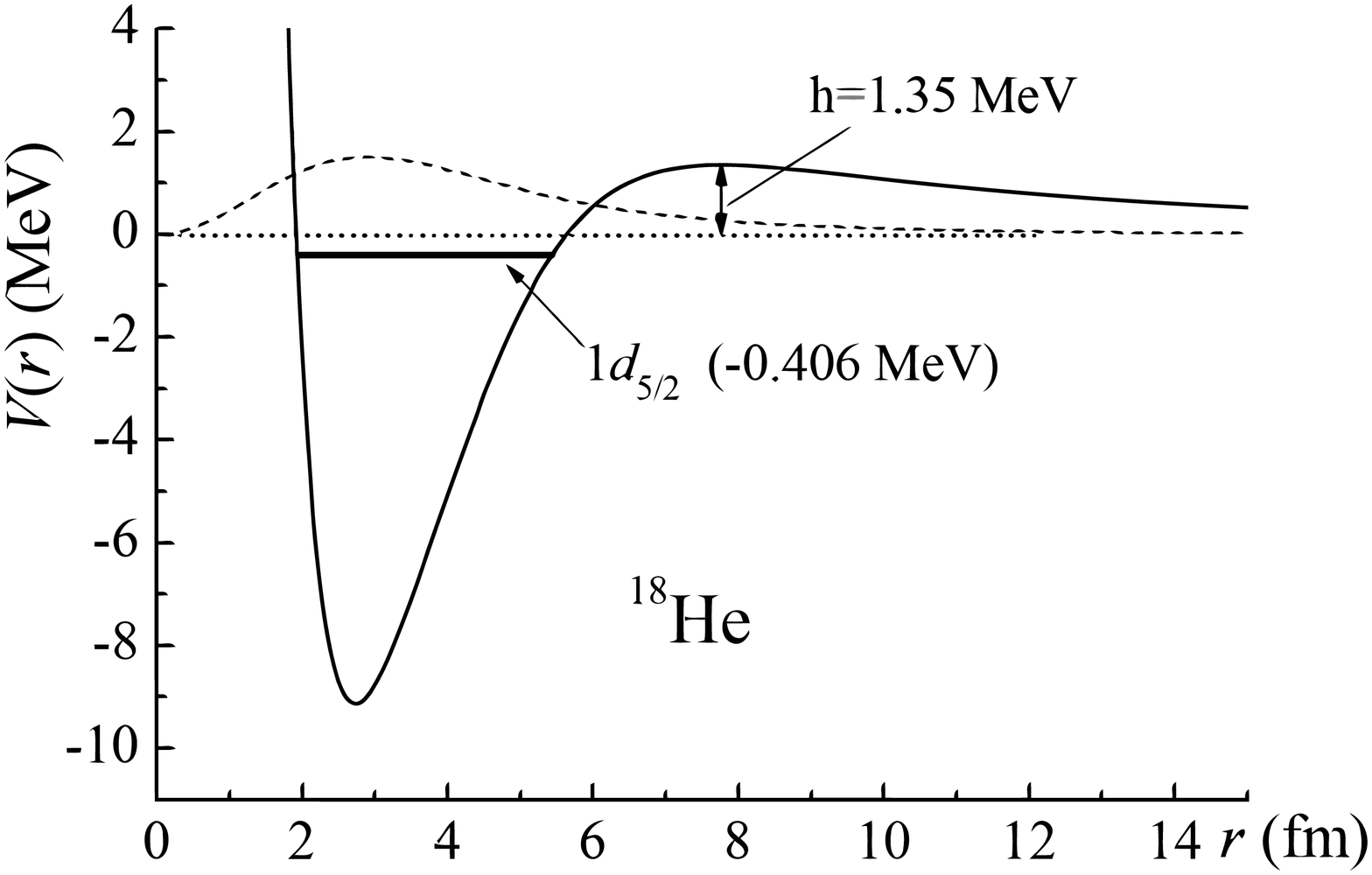}
}
\caption{The last filled level 1d$_{5/2}$ and the corresponding SHF potential $ 
V(r) $ for $^{18}$He (SkI2 forces). The energy is –0.406 MeV, the centrifugal 
barrier has the height 1.35 MeV. The dotted line is the wave function of the 
level 1d$_{5/2}$. }
\end{figure}

\begin{figure}[th]
\centering
\resizebox{1\textwidth}{!}{%
  \includegraphics{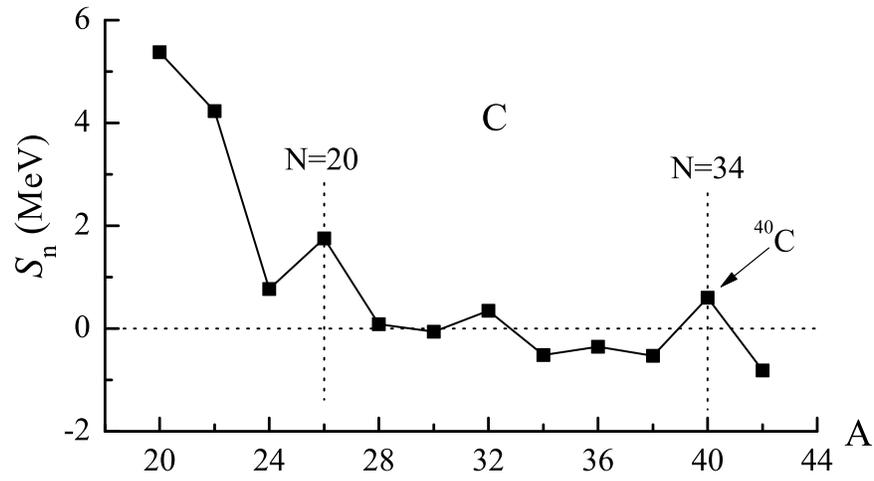}
}
\caption{The same as Fig. 2 but for carbon isotopes. The arrow points at   
$^{40}$C. }
\end{figure}

\begin{figure}[th]
\centering
\resizebox{1\textwidth}{!}{%
  \includegraphics{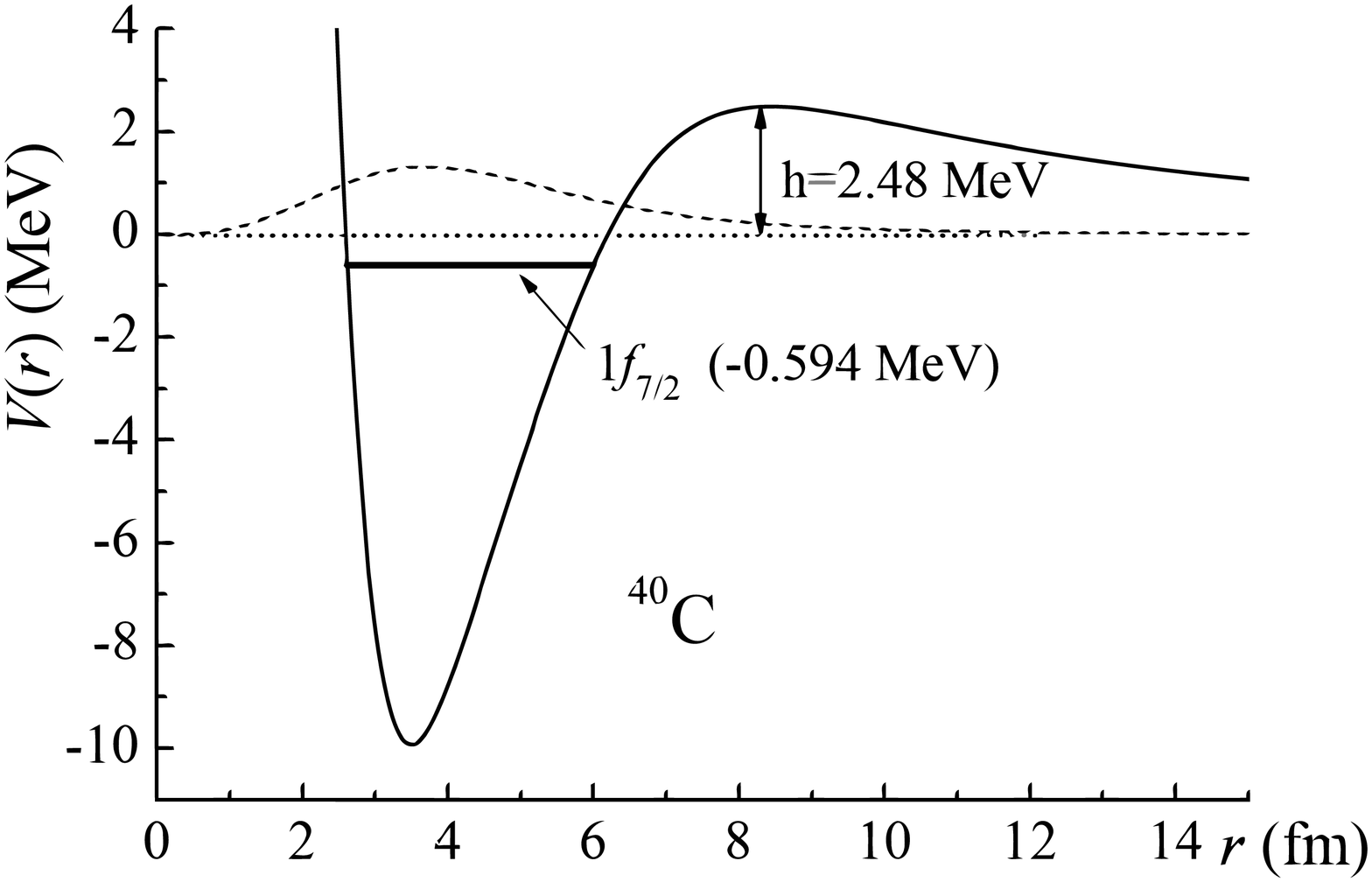}
}
\caption{The last filled level 1f$_{7/2}$ and the corresponding SHF potential $ 
V(r) $ for $^{40}$C (SkI2 forces). The energy is –0.594 MeV, the centrifugal 
barrier has the height 2.48 MeV. The dotted line is the wave function of the 
level 1f$_{7/2}$. }
\end{figure}

\begin{figure}[th]
\centering
\resizebox{1\textwidth}{!}{%
  \includegraphics{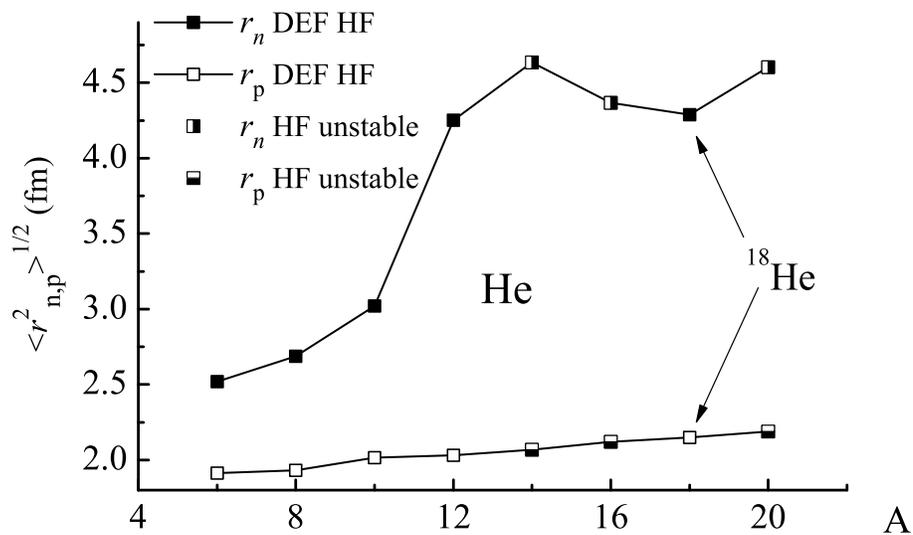}
}
\caption{ Neutron (filled squares) and proton (empty squares) root mean square 
radii of helium isotopes as functions of mass number. These are results of DHF 
calculations with SkI2 forces. Half-filled squares correspond to nuclei that are 
unstable against one neutron emission. }
\end{figure}

\begin{figure}[th]
\centering
\resizebox{1\textwidth}{!}{%
  \includegraphics{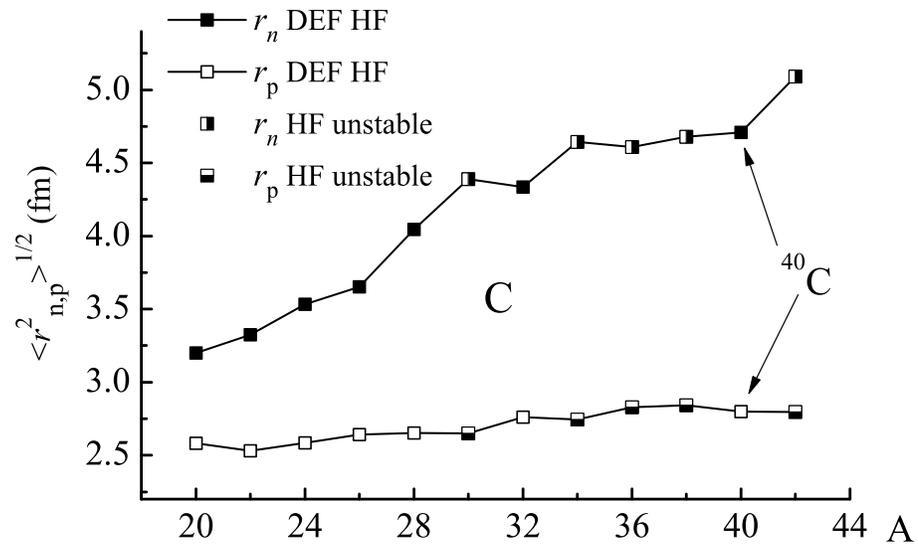}
}
\caption{The same as Fig. 6 but for carbon isotopes.  }
\end{figure}

\begin{figure}[th]
\centering
\resizebox{1\textwidth}{!}{%
  \includegraphics{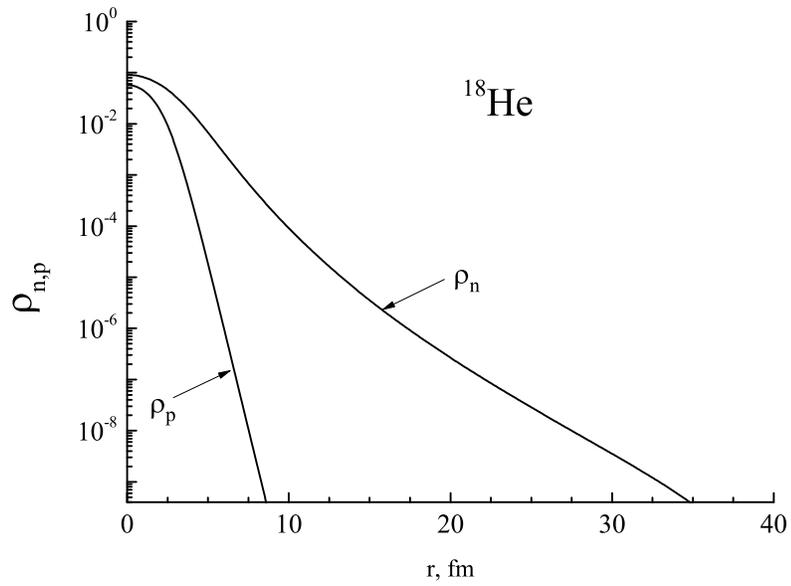}
}
\caption{ Neutron and proton density distributions for $^{18}$He calculated 
within SHF approach with SkI2 forces. }
\end{figure}

\begin{figure}[th]
\centering
\resizebox{1\textwidth}{!}{%
  \includegraphics{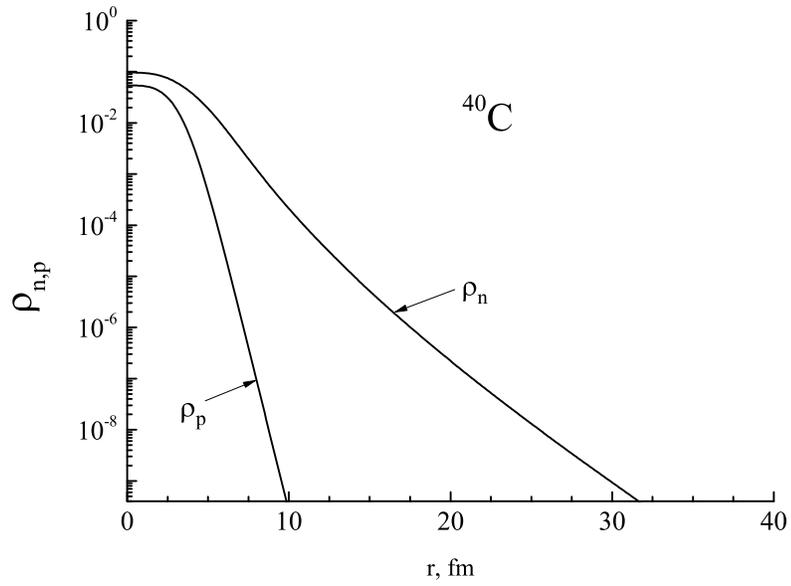}
}
\caption{The same as Fig. 8 but for $^{40}$C. }
\end{figure}

\end{document}